\documentclass[a4paper,reqno]{amsart}
\usepackage[all]{xy} 
\usepackage{amssymb} 
\usepackage{hyperref}
\usepackage{eucal}
\numberwithin{equation}{section}

\newtheorem{definition}{Definition}[section]

\newtheorem{theorem}[definition]{Theorem}

\newtheorem{remarkth}[definition]{Remark}
\newtheorem{example}[definition]{Example}
\newenvironment{remark}{\begin{remarkth}\upshape}{\hfill$\diamond$\end{remarkth}}

\renewcommand{\emph}[1]{{\bfseries\itshape{#1}}}

\newcommand{\R}{\mathbb{R}} 

\makeatletter
\newcommand\prol{\@ifstar{\@proldf}{\@prolpf}} 
\def\@prolpf{\@ifnextchar[{\@prolpf@wrt}{\@prolpf@}}
\def\@prolpf@wrt[#1]#2{\@ifnextchar[{\@prolpf@wrt@at{#1}{#2}}{\@prolpf@wrt@{#1}{#2}}}

\def\@prolpf@wrt@at#1#2[#3]{\prolsymbol^{#1}_{#3}#2}
\def\@prolpf@wrt@#1#2{\prolsymbol^{#1}#2}
\def\@prolpf@#1{\@ifnextchar[{\@prolpf@at{#1}}{\@prolpf@@{#1}}}
\def\@prolpf@at#1[#2]{\prolsymbol_{#2}#1}
\def\@prolpf@@#1{\prolsymbol#1}
\def\@proldf{\@ifnextchar[{\@proldf@wrt}{\@proldf@}}
\def\@proldf@wrt[#1]#2{\@ifnextchar[{\@proldf@wrt@at{#1}{#2}}{\@proldf@wrt@{#1}{#2}}}

\def\@proldf@wrt@at#1#2[#3]{\prolsymbol^{*#1}_{#3}#2}
\def\@proldf@wrt@#1#2{\prolsymbol^{*#1}#2}
\def\@proldf@#1{\@ifnextchar[{\@proldf@at{#1}}{\@proldf@@{#1}}}
\def\@proldf@at#1[#2]{\prolsymbol^*_{#2}#1}
\def\@proldf@@#1{\prolsymbol^*#1}
\def\prolsymbol{\mathcal{T}}
\makeatother

\begin{document}

\title[ Hamilton-Jacobi Theory for Nonholonomic Mechanical
Systems]{Towards a Hamilton-Jacobi Theory for Nonholonomic
Mechanical Systems}

\author[D.\ Iglesias-Ponte]{David Iglesias-Ponte}
\address{D.\ Iglesias-Ponte:
Instituto de Matem\'aticas y F{\'\i}sica Fundamental, Consejo
Superior de Investigaciones Cient{\'\i}ficas, Serrano 123, 28006
Madrid, Spain} \email{iglesias@imaff.cfmac.csic.es}

\author[M. de Le\'on]{Manuel de Le\'on}
\address{M. de Le\'on: Instituto de Matem\'aticas y F{\'\i}sica
Fundamental, Consejo
Superior de Investigaciones Cient{\'\i}ficas, Serrano 123, 28006
Madrid, Spain} \email{mdeleon@imaff.cfmac.csic.es}

\author[D.\ Mart{\'\i}n de Diego]{David Mart{\'\i}n de Diego}
\address{D.\ Mart{\'\i}n de Diego:
Instituto de Matem\'aticas y F{\'\i}sica Fundamental, Consejo
Superior de Investigaciones Cient{\'\i}ficas, Serrano 123, 28006
Madrid, Spain} \email{d.martin@imaff.cfmac.csic.es}



\begin{abstract}
In this paper we obtain a Hamilton-Jacobi theory for nonholonomic
mechanical systems. The results are applied to a large class of
nonholonomic mechanical systems, the so-called \v{C}aplygin systems.
\end{abstract}

\thanks{This work has been partially supported by MEC (Spain)
Grants MTM 2004-7832, S-0505/ESP/0158 of the CAM and ``Ingenio
Mathematica'' (i-MATH) No. CSD 2006-00032 (Consolider-Ingenio
2010). D. Iglesias acknowledges Ministry of Education and Science
for a ``Juan de la Cierva" research contract.}

\maketitle

\section{Introduction}

The standard formulation of the Hamilton-Jacobi problem for a
hamiltonian system is look for a function $S(t, q^A)$ (called
the {\bf principal function}) such that
\begin{equation}\label{hj1}
\frac{\partial S}{\partial t} + H(q^A, \frac{\partial S}{\partial
q^A}) = 0,
\end{equation}
where $H: T^*Q\longrightarrow \R$ is the hamiltonian function. If
one looks for solutions of the form $S(t, q^A) = W(q^A) - t E$,
where $E$ is a constant, then $W$ must satisfy
\begin{equation}\label{hj2}
H(q^A, \frac{\partial W}{\partial q^A}) = E,
\end{equation}
where $W$ is called the {\bf characteristic function}.

Equations (\ref{hj1}) and (\ref{hj2}) are indistinctly referred as
the {\bf Hamilton-Jacobi equation}.

The powerful of this method is that, in spite of the difficulties
to solve a partial differential equation instead of an ordinary
differential one, in many cases it works, being an extremely
useful tool, usually more than Hamilton's equations. Indeed, in
these cases the method provides an immediate way to integrate the
equations of motion. The modern interpretation relating the
Hamilton-Jacobi procedure with the theory of lagrangian
submanifolds is an important source of new results and insights
\cite{AM,arnold}. Let us remark that, recently, Cari{\~{n}}ena
{\it et al} \cite{CGMMMR} have developed a new approach to the
geometric Hamilton-Jacobi theory.

On the other hand, in the last fifteen years there has been a
renewed interest in nonholonomic mechanics, that is, those
mechanical systems given by a lagrangian $L=L(q^A, \dot{q}^A)$
subject to constraints $\Phi^i(q^A, \dot{q}^A) = 0$ involving the
velocities (see \cite{B} and references therein). A relevant
difference with the unconstrained mechanical systems is that a
nonholonomic system is not hamiltonian in the sense that the phase
space is just the constraint submanifold and not the cotangent
bundle of the configuration manifold; moreover, its dynamics is
given by an almost Poisson bracket, that is, a bracket not
satisfying the Jacobi identity \cite{CaLeMa}. In \cite{MdLDMdD0}
the authors proved that the nonholonomic dynamics can be obtained
by projecting the unconstrained dynamics; this will be the point
of view adopted in the present paper.

A natural question, related with a possible notion of
integrability is in what extent one could construct a
Hamilton-Jacobi theory for nonholonomic mechanics. Past attempts
to obtain a Hamilton-Jacoby theory for nonholonomic systems were
non-effective or very restrictive (see
\cite{Eden,Doo1,Doo2,Doo3,Doo4} and also \cite{Pa}), because, in
many of them, they try to adapt the typical proof of the
Hamilton-Jacobi equations for systems without constraints (using
Hamilton's principle). Usually the results are valid when the
solutions of the nonholonomic problem are also the solutions of
the corresponding constrained variational problem (see
\cite{Kozlov,Rumyantsev,Sumbatov} for a complete discussion).

In our paper, we present an alternative approach based on the
geometrical properties of nonholonomic systems (see also
\cite{Prince} for second-order differential equations). The method
is applied to a particular class of nonholonomic systems,
called \v{C}aplygin systems: in such a 
system the configuration manifold is a fibration over another
manifold, and the constraints are given by the horizontal
subspaces of a connection on the fibration. In this case, the
original nonholonomic system is equivalent to another one whose
configuration manifold is the base of the fibration and, in
addition, it is subject to an external force \cite{MdLDMdD1}. In
any case, the equations we obtained are different that in previous
works and may give new insight in this topic. In particular, this
theory could give insights in the study of integrability for
nonholonomic systems \cite{larry1} and even in the construction of
new geometrical integrators for nonholonomic systems (see
\cite{Hairer,SaLeMa}).

\section{Preliminaries}

\subsection{Lagrangian and Hamiltonian mechanics}

Let $L=L(q^A, \dot{q}^A)$ be a lagrangian function, where $(q^A)$
are coordinates in a configuration $n$-manifold $Q$. Hamilton's
principle produces the Euler-Lagrange equations
\begin{equation}\label{eleqs}
\frac{d}{dt}\left(\frac{\partial L}{\partial \dot{q}^A}\right) -
\frac{\partial L}{\partial q^A} = 0, \; 1 \leq A \leq n.
\end{equation}

A geometric version of Eq. (\ref{eleqs}) (see \cite{MdLPR}) can be
obtained as follows. Consider the (1,1)-tensor field $S$ and the
Liouville vector field $\Delta$ locally defined on the tangent
bundle $TQ$ of $Q$ by
$$
S = \frac{\partial }{\partial \dot{q}^A} \otimes d q^A, \; \quad
\Delta = \dot{q}^A \frac{\partial }{\partial \dot{q}^A}.
$$
Since the lagrangian $L$ is a function defined on $TQ$ one can
construct the Poincar\'e-Cartan 1- and 2-forms
$$
\alpha_L = S^*(dL), \; \quad \omega_L = - d \alpha_L,
$$
where $S^*$ denotes the adjoint operator of $S$. The energy is
given by $E_L = \Delta (L) - L.$ We say that $L$ is regular if the
2-form $\omega_L$ is symplectic. In this case, the equation
\begin{equation}\label{seleqs}
i_X \, \omega_L = dE_L
\end{equation}
has a unique solution, $X=\xi_L$, called the Euler-Lagrange vector
field; $\xi_L$ is a second order differential equation (SODE)
which means that its integral curves are tangent lifts of their
projections on $Q$ (these projections are called the solutions of
$\xi_L$). A direct computation shows that the solutions of $\xi_L$
are just the ones of Eqs. (\ref{eleqs}).

Finally, let us recall that the Legendre transformation $FL : TQ
\longrightarrow T^*Q$ is a fibred mapping (that is, $\pi_Q \circ
FL = \tau_Q$, where $\tau_Q : TQ \longrightarrow Q$ and $\pi_Q :
T^*Q \longrightarrow Q$ denote the canonical projections of the
tangent and cotangent bundles of $Q$, respectively). The
regularity of $L$ is equivalent to $FL$ being a local
diffeomorphism. Along this paper, we will assume that $FL$ is in
fact a global diffeomorphism (in other words, $L$ is hyperregular)
which is the case when $L$ is a lagrangian of mechanical type, say
$L=T-V$, where $T$ is the kinetic energy defined by a Riemannian
metric on $Q$ and $V : Q \longrightarrow \R$ is a potential
energy.

The hamiltonian counterpart is developed in the cotangent bundle
$T^*Q$ of $Q$. Denote by $\omega_Q = dq^A \wedge dp_A$ the
canonical symplectic form, where $(q^A, p_A)$ are the canonical
coordinates on $T^*Q$. The Hamiltonian energy is just $H = E_L
\circ FL^{-1}$ and the Hamiltonian vector field is the solution of
the symplectic equation
$$
i_{X_H} \, \omega_Q = dH.
$$
As we know, the integral curves $(q^A(t), p_A(t))$ of $X_H$
satisfy the Hamilton equations
\begin{equation}\label{hamiltoneqs}
\left.
\begin{array}{lcr}
\dot{q}^A & = &\displaystyle{ \frac{\partial H}{\partial
p_A}}\\[7pt]
\dot{p}_A &=&\displaystyle{ - \frac{\partial H}{\partial q^A}}
\end{array}
\right\}
\end{equation}

Finally, since $FL^* \omega_Q = \omega_L$ we deduce that $\xi_L$
and $X_H$ are $FL$-related and, consequently, $FL$ transforms the
Euler-Lagrange equations (\ref{eleqs}) into the Hamilton equations
(\ref{hamiltoneqs}).

\subsection{Nonholonomic mechanical systems}

A nonholonomic mechanical system is given by a lagrangian function
$L = L(q^A, \dot{q}^A)$ subject to a family of constraint
functions
$$
\Phi^i (q^A, \dot{q}^A) = 0, \; 1 \leq i \leq m \leq n = \dim Q.
$$
In the sequel, we will assume that the constraints $\Phi^i$ are
linear in the velocities, i.e., $\Phi^i (q^A, \dot{q}^A) =
\Phi^i_A(q) \dot{q}^A$.

Invoking the D'Alembert principle, we derive the nonholonomic
equations of motion
\begin{equation}\label{nheqs}
\left.
\begin{array}{rcl}
\displaystyle{
\frac{d}{dt}\left(\frac{\partial L}{\partial
\dot{q}^A}\right) - \frac{\partial L}{\partial q^A}} & =
&\lambda_i \Phi^i_A(q)\; ,\qquad 1 \leq A \leq n\\
\Phi^i(q^A, \dot{q}^A) & = &0\; ,\qquad 1 \leq i \leq m
\end{array}
\right\}
\end{equation}
where $\lambda_i = \lambda_i (q^A, \dot{q}^A)$, $1 \leq i \leq m$,
are Lagrange multipliers to be determined.

In a geometrical setting, $L$ is a function on $TQ$ and the
constraints are given by a vector subbundle $M$ of $TQ$ locally
defined by $\Phi^i = 0$.

Equations (\ref{nheqs}) can be intrinsically (see
\cite{MdLDMdD1}) rewritten as follows
\begin{equation}\label{nheqs3}
\left.
\begin{array}{rcl}
i_X \, \omega_L - dE_L & \in & S^*((TM)^0)\\
X & \in & TM.
\end{array}
\right\}
\end{equation}

For the formulation of a Hamilton-Jacobi theory we are interested
in the ``Hamiltonian version" of the nonholonomic equations.
Assuming that the Lagrangian $L$ is hyperregular, then the
constraint functions on $T^*Q$ become $\Psi^i=\Phi^i\circ
FL^{-1}$, i.e.
\[
\Psi^i(q^A, p_A)=\Phi^i_A(q)\frac{\partial H}{\partial p_A}(q^A,
p_A)\,,
\]
where the Hamiltonian $H: T^*Q\rightarrow \R$ is defined by $H=E_L
\circ FL^{-1}$.

The equations of motion for the nonholonomic system on $T^*Q$ can
now be written as follows
\begin{equation}\label{hnh}
\left.
\begin{array}{rcl}
\dot q^A&=&\displaystyle{\frac{\partial H}{\partial p_A}}\\
\vphantom{\huge A}\dot p_A&=&\displaystyle{-\frac{\partial
H}{\partial q^A}-\bar{\lambda}_i \Phi^i_A(q)}
\end{array}\right\}
\end{equation}
together with the constraint equations $\Psi^i(q,p)=0$.

 Let
$\bar{M}$ denote the image of the constraint submanifold $M$ under
the Legendre transformation, and let $\bar{F}$ be the distribution
on $T^*Q$ along $\bar{M}$, whose annihilator is given by
\[
\bar{F}^0 = FL_*( S^*((T{M})^0))\,.
\]
Observe that $\bar{F}^0$ is locally generated by the $m$ independent
1-forms
\[
\bar{\mu}^i=\Phi^i_A(q)d q^A\; ,\ 1\leq i\leq m \,.
\]
The nonholonomic Hamilton equations for the nonholonomic system
can be then rewritten in intrinsic form as
\begin{equation}\label{a1}
\left.
\begin{array}{rcl}
(i_X\omega_Q-dH)_{|\bar{M}}&\in& \bar{F}^{0}\\
X_{|\bar{M}} &\in& T\bar{M}
\end{array}
\right\}
\end{equation}

Assume the compatibility condition: $\bar{F}^{\perp}\cap
T\bar{M}=\{0\}$, where $``\perp"$ denotes the symplectic
orthogonal with respect to $\omega_Q$. Observe that, locally, this
condition means that the matrix
\begin{equation}\label{a2}
(\bar{\mathcal{C}}^{ij})= \left(\Phi^i_A(q){\mathcal
H}^{AB}\Phi^j_B(q)\right)
\end{equation}
is regular, where $({\mathcal H}^{AB})=(\partial^2 H/ \partial
p_A\partial p_B)$. The compatibility condition is not too
restrictive, since it is trivially verified by the usual systems
of mechanical type (i.e.\ with a Lagrangian of the form kinetic
minus potential energy). The compatibility condition guarantees in
particular the existence of a unique solution of the constrained
equations of motion (\ref{a1}) which, henceforth, will be denoted
by $\bar{X}_{nh}$.

Moreover, if we denote by $X_H$ the Hamiltonian vector field of $H$,
i.e. $i_{X_H}\omega_Q=dH$ then, using the constraint functions, we
may explicitly determine the Lagrange multipliers $\lambda_i$ as
\begin{equation}\label{qer}
\bar{\lambda}_i= \bar{\mathcal C}_{ij} X_H(\Psi^j)\; ,
\end{equation}
where $(\bar{\mathcal C}_{ij})$ is the inverse matrix of
$(\bar{\mathcal C}^{ij})$.

\subsection{\v{C}aplygin systems}\label{chaply}

A \v{C}aplygin system is a nonholonomic mechanical system such that:
\begin{enumerate}
\item the configuration manifold $Q$ is a fibred manifold, say $\rho : Q
\longrightarrow N$, over a manifold $N$;
\item the constraints are provided by the horizontal distribution of
an Ehresmann connection $\Gamma$ in $\rho$;
\item the lagrangian $L : TQ \longrightarrow \R$ is
$\Gamma$-invariant.
\end{enumerate}
\begin{remark}
A particular case is when $\rho : Q \longrightarrow N=Q/G$ is a
principal $G$-bundle and $\Gamma$ a principal connection.
\end{remark}
Let us recall that the connection $\Gamma$ induces a Whitney
decomposition $ TQ = {\mathcal H} \oplus V\rho $ where
$\mathcal{H}$ is the horizontal distribution, and $V\rho = \ker
T\rho$ is the vertical distribution. Take fibred coordinates
$(q^A) = (q^a, q^i)$ such that $\rho (q^a, q^i) = (q^a)$;
therefore we can obtain an adapted local basis of vector fields
\[
{\mathcal H} = \langle {\mathcal H}_a = \frac{\partial}{\partial
q^a} - \Gamma^i_a \frac{\partial}{\partial q^i} \rangle, \qquad
V\rho = \langle V_i = \frac{\partial}{\partial q^i} \rangle.
\]
Here $\displaystyle{{\mathcal H}_a = (\frac{\partial}{\partial
q^a})^{\mathcal H} = h (\frac{\partial}{\partial q^a})}$, where
$y^{\mathcal H}$ denotes the horizontal lift of a tangent vector
$y$ on $N$ to $Q$, and $h: TQ \longrightarrow {\mathcal H}$ is the
horizontal projector; $\Gamma^i_a = \Gamma^i_a (q^A)$ are the
Christoffel components of the connection $\Gamma$.

The dual local basis of 1-forms is
$$
\{\eta_a = dq^a, \eta_i = dq^i + \Gamma^i_a dq^a\}
$$

The curvature of $\Gamma$ is the (1,2)-tensor field
$
R = \frac{1}{2} [h, h]
$
where $[\, , \, ]$ is the Nijenhuis tensor of $h$, that is
$$
R(X, Y) = [hX, hY] - h[hX, Y] - h[X, hY] + h^2[X, Y]
$$
Therefore we have
$$
R(\frac{\partial}{\partial q^a}, \frac{\partial}{\partial q^b}) =
R^i_{ab} \frac{\partial}{\partial q^i}
$$
where
$$
R^i_{ab} = \frac{\partial \Gamma^i_a}{\partial q^b} - \frac{\partial
\Gamma^i_b}{\partial q^a} + \Gamma^j_a \frac{\partial
\Gamma^i_b}{\partial q^j} - \Gamma^j_b \frac{\partial
\Gamma^i_a}{\partial q^j}
$$
The constraints are locally given by $\Phi^i = \dot{q}^i +
\Gamma^i_a \dot{q}^a = 0.$ In other words, the solutions are
horizontal curves with respect to $\Gamma$.

Since the lagrangian $L$ is $\Gamma$-invariant, that is, $
L((Y^{\mathcal H})_{q_1}) = L((Y^{\mathcal H})_{q_2}) $ for all $Y
\in T_yN$, $y = \rho(q_1) = \rho(q_2)$, we can define a function
$L^* : TN \longrightarrow \R$ as follows: $ L^* (Y_y) =
L((Y^{\mathcal H})_q)$, where $y=\rho(q)$. Therefore we have
$$
L^*(q^a, \dot{q}^a) = L(q^a, q^i, \dot{q}^a, -\Gamma^i_a \dot{q}^a)
$$

Equations (\ref{nheqs3}) read now as
\begin{equation}\label{nheqs4}
\left.
\begin{array}{rcl}
i_X \, \omega_L - dE_L & \in & S^*((T{\mathcal H})^0)\\
X & \in & T{\mathcal H}
\end{array}
\right\}
\end{equation}

Define a 1-form $\alpha^*$ on $TN$ by putting
$$
(\alpha^*)(u)(U) = - (\alpha_L)(x) (\tilde{u}),
$$
where $U \in T_u (TN)$, $u \in T_yN$, $\tilde{U} \in T_x(TQ)$ such
that $\tilde{U}$ projects onto $$R((u)^{\mathcal H}_q,
(T\tau_N(U)^{\mathcal H}_q)) \in T_qQ,$$ $\rho(q) = y$, $x \in
{\mathcal H}$, $\tau_Q(x) =q$. In local coordinates we obtain
$$
\alpha^* = \left(\frac{\partial L}{\partial \dot{q}^i} \dot{q}^b
R^i_{ab}\right) \, dq^a.
$$

Consider the following equation
\begin{equation}\label{eqs*}
i_Y \, \omega_{L^*} - dE_{L^*} = \alpha^*.
\end{equation}

A long but straightforward proof shows that $L^*$ is a regular
lagrangian on $TN$, therefore (\ref{eqs*}) has a unique solution
$Y^*$. Notice that the pair $(L^*, \alpha^*)$ can be considered as
an unconstrained system subject to an external force $\alpha^*$.
The corresponding equations of motion are
\begin{equation}\label{eleqs*}
\frac{d}{dt}\left(\frac{\partial L^*}{\partial \dot{q}^a}\right) -
\frac{\partial L^*}{\partial q^a} = - \frac{\partial L}{\partial
\dot{q}^i} \dot{q}^b R^i_{ab}.
\end{equation}

Both systems, the nonholonomic one on $Q$ given by $L$ and the
constraints given by $\Gamma$, and that given by $L^*$ and
$\alpha^*$, are equivalent. The equivalence is explained in the
following.

$\Gamma$ induces a connection $\bar{\Gamma}$ in the fibred
manifold $T\rho : TQ \longrightarrow TN$ along ${\mathcal H}$ by
defining its horizontal distribution as follows:

\begin{eqnarray*}
(\frac{\partial}{\partial q^a})^{\bar{{\mathcal H}}} & = &
\frac{\partial}{\partial q^a} - \Gamma^i_a
\frac{\partial}{\partial q^i} - \left(\dot{q}^b \frac{\partial
\Gamma^i_b}{\partial q^a} - \Gamma^j_a \frac{\partial
\Gamma^i_b}{\partial q^j}\right)
\frac{\partial}{\partial \dot{q}^i}\\
(\frac{\partial}{\partial \dot{q}^a})^{\bar{{\mathcal H}}} & = &
\frac{\partial}{\partial \dot{q}^a} - \Gamma^i_a
\frac{\partial}{\partial \dot{q}^i}
\end{eqnarray*}

\begin{theorem}
The nonholonomic dynamics $X_{nh}$ is a vector field on ${\mathcal
H}$ which is $T\rho$-projectable onto $Y^*$. Furthermore, $X_{nh}$
is the horizontal lift of $Y^*$ with respect to the induced
connection $\bar{\Gamma}$.
\end{theorem}

\begin{example} (Mobile robot with fixed orientation)
{\rm

The body of the robot maintains a fixed orientation with respect
to the environment. The robot has three wheels with radius $R$,
which turn simultaneously about independent axes, and perform a
rolling without sliding over a horizontal floor.

Let $(x, y)$ denotes the position of the centre of mass, $\theta$
the steering angle of the wheel, $\psi$ the rotation angle of the
wheels in their rolling motion over the floor. So, the
configuration manifold is $Q = S^1 \times S^1 \times \R^2. $ The
lagrangian $L$ is
$$
L = \frac{1}{2} m \dot{x}^2 + \frac{1}{2} m \dot{y}^2 + \frac{1}{2}
J \dot{\theta}^2 + \frac{3}{2} J_\omega \dot{\psi}^2
$$
where $m$ is the mass, $J$ is the moment of inertia and $J_\omega$
is the axial moment of inertia of the robot.

The constraints are induced by the conditions that the wheels roll
without sliding, in the direction in which they point, and that the
instantaneous contact point of the wheels with the floor have no
velocity component orthogonal to that direction:
\begin{eqnarray*}
\dot{x} \sin \theta - \dot{y} \cos \theta & = & 0, \\
\dot{x} \cos \theta + \dot{y} \sin \theta - R \dot{\psi} & = & 0.
\end{eqnarray*}

The abelian group $G = \R^2$ acts on $Q$ by translations, say
$$
((a, b), (\theta, \psi, x, y)) \mapsto (\theta, \psi, a+x, b+y)
$$
Therefore we have a principal $G$-bundle $\rho: Q \longrightarrow N
= Q/G$ with a principal connection given by the connection 1-form
$$
\beta = (dx - R \cos \theta d\psi) e_1 + (dy - R \sin \theta d\psi)
e_2
$$
where $\{e_1, e_2\}$ denotes the standard basis of $\R^2$. The
constraints are given by the horizontal subspaces of $\beta$. If we
apply the above reduction procedure we deduce $\alpha^* = 0$.
}
\end{example}

\section{Geometric Hamilton-Jacobi theory}

The following result is a geometric version of the standard
formulation of the Hamilton-Jacobi problem \cite{AM}.

\begin{theorem}\label{AM}
Let $\gamma$ be a closed 1-form on $Q$. Then the following
conditions are equivalent:

\begin{enumerate}
\item[(i)] for every curve $\sigma: \R \longrightarrow Q$ such that
$$
\dot{\sigma}(t) = T\pi_Q(X_H(\gamma(\sigma(t))))
$$
for all $t$, then $\gamma \circ \sigma$ is an integral curve of
$X_H$.

\item[(ii)] $d(H \circ \gamma) = 0$.
\end{enumerate}
\end{theorem}

If $\gamma = dW$ we recover the standard formulation since $d(H
\circ dW) = 0$ is equivalent to the condition $H \circ dW = cte$,
that is
$$
H(q^A, \frac{\partial W}{\partial q^A}) = E
$$
where $E$ is a constant.

A interesting new point of view of the geometric Hamilton-Jacobi
theory has been recently developed by J.F. Cari\~{n}ena {\it et
al.} \cite{CGMMMR}.

Let $\gamma$ be a closed 1-form as in Theorem \ref{AM}. Since $FL$
is a diffeomorphism, we can define a vector field $X$ on $Q$ by
$$
X = FL^{-1} \circ \gamma
$$
Therefore, we have
$$
0 = d(H \circ \gamma) = d(E_L \circ FL^{-1} \circ \gamma) = d(E_L
\circ X)
$$
because $H = E_L \circ FL^{-1}$.

Hence, Theorem \ref{AM} can be reformulated as follows.

\begin{theorem}\label{AMvf}\cite{CGMMMR}
Let $X$ be a vector field on $Q$ such that $FL \circ X$ is a closed
1-form. Then the following conditions are equivalent:

\begin{enumerate}
\item[(i)] for every curve $\sigma: \R \longrightarrow Q$ such that
$$
\dot{\sigma}(t) = T\tau_Q(\xi_L(X(\sigma(t))))
$$
for all $t$, then $X \circ \sigma$ is an integral curve of $\xi_L$.

\item[(ii)] $d(E_L \circ X) = 0$.
\end{enumerate}
\end{theorem}

\begin{definition}
A vector field $X$ satisfying the conditions of Theorem \ref{AMvf}
will be called a solution for the Hamilton-Jacobi problem given by
$L$.
\end{definition}

\subsection{An interlude: mechanical systems with external forces}

We shall need the following formulation of the Hamilton-Jacobi
theory for mechanical systems with external forces.

A mechanical system with an external force is given by (see
\cite{Godb}):

\begin{enumerate}
\item A lagrangian function $L : TQ \longrightarrow \R$, where
$Q$ is the configuration manifold;

\item a semibasic 1-form $\alpha$ on $TQ$.

\end{enumerate}

Since $\alpha$ is semibasic (that means that $\alpha$ vanishes when
it is applied to vertical tangent vectors) we have
$$
\alpha = \alpha_A(q, \dot{q}) \, dq^A
$$

The Euler-Lagrange equations are then
\begin{equation}\label{exteleqs}
\frac{d}{dt}(\frac{\partial L}{\partial \dot{q}^A}) - \frac{\partial
L}{\partial q^A} = - \alpha_A, \; 1 \leq A \leq n,
\end{equation}
which correspond to the symplectic equation
\begin{equation}\label{sexteleqs}
i_X \, \omega_L = dE_L + \alpha
\end{equation}
Indeed, when $L$ is regular, Eq. (\ref{sexteleqs}) has a unique
solution $\xi_{L,\alpha}$ which is a second order differential
equation whose solutions are just the ones of (\ref{exteleqs}).

Notice that $\xi_{L,\alpha} = \xi_L + Z$, where $i_Z\omega_L=
\alpha$.

Observe that we can construct the hamiltonian counterpart using the
Legendre transformation, so that we have a hamiltonian $H = E_L
\circ FL^{-1}$ subject to the external force $\beta = (FL^{-1})^*
\alpha$ which is again semibasic (i.e. $\beta = \beta_A \, dq^A$).
The equation
$$
i_{X_{H,\beta}} \, \omega_Q = dH + \beta
$$
has a unique solution $X_{H,\beta}$ whose integral curves satisfy
the Hamilton equations with external force
\begin{equation}\label{hnhexternal}
\left.
\begin{array}{rcl}
\dot q^A&=&\displaystyle{\frac{\partial H}{\partial p_A}}\\
\vphantom{\huge A}\dot p_A&=&\displaystyle{-\frac{\partial
H}{\partial q^A}-\beta_A}
\end{array}\right\}
\end{equation}

\begin{theorem}\label{extAMvfe}
Let $\gamma$ be a closed 1-form on $Q$. Then the following
conditions are equivalent:

\begin{itemize}
\item [(i)] for every curve $\sigma: \R \longrightarrow Q$ such that
\begin{equation}\label{auxiliar}
\dot{\sigma}(t) = T\pi_Q(X_{H,\beta}(\gamma(\sigma(t))))
\end{equation}
for all $t$, then $\gamma \circ \sigma$ is an integral curve of
$X_{H,\beta}$.

\item [(ii)] $d(H \circ \gamma) = - \gamma^*\beta$.
\end{itemize}
\end{theorem}

\begin{proof} Since $\gamma=\gamma_A\, dq^A$ is closed then
\[
\frac{\partial \gamma_A}{\partial q^B}=\frac{\partial
\gamma_B}{\partial q^A}
\]
It is easy to show that Equation (\ref{auxiliar}) is rewritten, in
local coordinates, as
\begin{equation}\label{unodos0}
\dot{\sigma}^A(t)=\frac{\partial H}{\partial p_A}(\sigma^B(t),
\gamma_B(\sigma(t)))
\end{equation}

We also have that condition
\[
d(H \circ \gamma) = - \gamma^*\beta
\]
is written in local coordinates as
\begin{equation}\label{unotres0}
\frac{\partial H}{\partial q^A}+\frac{\partial H}{\partial
p_B}\frac{\partial \gamma_B}{\partial q^A}= - \beta_A
\end{equation}

($\Longrightarrow$) Assume that (i) holds. Therefore

\begin{equation}\label{aqqq0}
\frac{d}{dt}(\gamma_A(\sigma(t)))=-\frac{\partial H}{\partial
q^A}(\gamma(\sigma(t))) - \beta_A(\gamma(\sigma(t)))
\end{equation}

Moreover
\begin{eqnarray*}
\frac{\partial H}{\partial q^A}+\frac{\partial H}{\partial p_B}
\frac{\partial \gamma_B}{\partial q^A}&=&\frac{\partial H}{\partial
q^A} +\frac{\partial H}{\partial p_B}\frac{\partial
\gamma_A}{\partial q^B}
\hbox{ (since } \gamma \hbox{ is closed)}\\
&=& \frac{\partial H}{\partial q^A}+\dot{\sigma}^B(t)
\frac{\partial \gamma_A}{\partial q^B} \quad \hbox{(from
(\ref{unodos0}))} \\
&=&-\beta_A} \quad \hbox{(from (\ref{aqqq0}))
\end{eqnarray*}

($\Longleftarrow$) Assume that (ii) holds, that is,
\[
\frac{\partial H}{\partial q^A}+\frac{\partial H}{\partial
p_B}\frac{\partial \gamma_B}{\partial q^A} = - \beta_A
\]
Now using (\ref{unodos0}) and since $\gamma$ is closed, then
\[
\frac{\partial H}{\partial q^A}+\dot{\sigma}^B(t)\frac{\partial
\gamma_A}{\partial q^B}= - \beta_A
\]
Therefore
$$
\frac{d}{dt}(\gamma_A(\sigma(t))) = - \frac{\partial H}{\partial
q^A}(\gamma(\sigma(t))) - \beta_A(\gamma(\sigma(t)))
$$
which proves that $\gamma \circ \sigma$ is an integral curve of
$X_{H,\beta}$.
\end{proof}

Therefore we have the lagrangian version.

\begin{theorem}\label{extAMvfe1}
Let $X$ be a vector field on $Q$ such that $FL \circ X$ is a closed
1-form. Then the following conditions are equivalent:

\begin{itemize}
\item [(i)] for every curve $\sigma: \R \longrightarrow Q$ such that
$$
\dot{\sigma}(t) = T\tau_Q(\xi_{L,\alpha}(X(\sigma(t))))
$$
for all $t$, then $X \circ \sigma$ is an integral curve of
$\xi_{L,\alpha}$.

\item [(ii)] $d(E_L \circ X) = - X^*\alpha$.
\end{itemize}
\end{theorem}

\begin{definition}
A vector field $X$ satisfying the conditions of Theorem
\ref{extAMvfe1} will be called a solution for the Hamilton-Jacobi
problem given by $L$ and $\alpha$.
\end{definition}

\section{Hamilton-Jacobi theory for nonholonomic mechanical systems}

Let $L:TQ \longrightarrow \R$ be a lagrangian function subject to
nonholonomic constraints given by a vector subbundle $M$ of $TQ$,
locally defined by the linear constraints
$\Phi^i=\Phi^i_A(q)\dot{q}^A$, $1\leq i\leq m$. Denote by $D$ the
distribution on $Q$ whose annihilator is $D^0=\hbox{span} \{
\mu^i=\Phi^i_A(q)dq^A\}$. Notice that $S^*(TM^0)$ is the pullback
to $TQ$ of the annihilator $D^0$ of $D$.

We assume the admissibility and compatibility conditions, and
consider the hamiltonian counterpart given by a Hamiltonian
function $H¨: T^*Q \longrightarrow \R$ and a constraint
submanifold $\bar{M}=FL(M)$ as in the precedent sections. $X_{nh}$
and $\bar{X}_{nh}$ will denote the corresponding nonholonomic
dynamics. Given $D^0$, the annihilator of $D$,  we can form the
algebraic ideal ${\mathcal I} (D^0)$ in the algebra
$\Lambda^*(Q)$. Therefore, if a $k$-form $\nu\in {\mathcal I}
(D^0)$ then
\[
\nu=\beta_i\wedge \mu^i, \qquad \hbox{where}\qquad \beta_i\in
\Lambda^{k-1}(Q), \quad 1\leq i\leq m.
\]

\begin{theorem}\label{nhhj1}
Let $\gamma$ be a 1-form on $Q$ such that $\gamma(Q) \subset
\bar{M}$ and $d\gamma\in {\mathcal I} (D^0)$. Then the following
conditions are equivalent:

\begin{itemize}
\item[(i)] for every curve $\sigma: \R \longrightarrow Q$ such that
\begin{equation}\label{uno}
\dot{\sigma}(t) = T\pi_Q(X_H(\gamma(\sigma(t))))
\end{equation}
for all $t$, then $\gamma \circ \sigma$ is an integral curve of
$\bar{X}_{nh}$.

\item[(ii)] $d(H \circ \gamma) \in D^0$.
\end{itemize}
\end{theorem}
\begin{proof}
The condition $d\gamma\in {\mathcal I} (D^0)$ means that
\[
\frac{\partial \gamma_A}{\partial q^B}=\frac{\partial
\gamma_B}{\partial q^A}+\beta_{iA}\Phi^i_B-\beta_{iB}\Phi^i_A
\]
where $\gamma=\gamma_A dq^A$ and $\beta _i=\beta_{iA}dq^A$.
It is
easy to show that Equation (\ref{uno}) is rewritten, in local
coordinates, as
\begin{equation}\label{unodos}
\dot{\sigma}^A(t)=\frac{\partial H}{\partial p_A}(\sigma^B(t),
\gamma_B(\sigma(t)))
\end{equation}

We also have that condition
\[
d(H\circ \gamma)\in D^0
\]
is written in local coordinates as
\begin{equation}\label{unotres}
\left[\frac{\partial H}{\partial q^A}+\frac{\partial H}{\partial
p_B}\frac{\partial \gamma_B}{\partial q^A}\right]\,
dq^A=\tilde{\lambda}_i {\mu}^i=\tilde{\lambda}_i \Phi^i_A(q)\,
dq^A
\end{equation}
for some Lagrange multipliers $\tilde{\lambda}_i$'s.

($\Longrightarrow$) Assume that (i) holds. Therefore

\begin{equation}\label{aqqq}
\frac{d}{dt}(\gamma_A(\sigma(t)))=-\frac{\partial H}{\partial
q^A}(\gamma(\sigma(t))) - \bar{\lambda}_i \Phi^i_A(\sigma(t))\, ,
\end{equation}
where the $\bar{\lambda}_i$'s are determined using the constraint
equations \[ \Psi^i(\sigma(t),\gamma(\sigma(t)))=\frac{\partial
H}{\partial
p_B}(\sigma(t),\gamma(\sigma(t)))\Phi^i_B(\sigma(t))=0.
\]
Using the constraint equations we deduce that
\begin{eqnarray*}
\frac{\partial H}{\partial q^A}+\frac{\partial H}{\partial p_B}
\frac{\partial \gamma_B}{\partial q^A}&=&\frac{\partial
H}{\partial q^A} +\frac{\partial H}{\partial p_B}\frac{\partial
\gamma_A}{\partial q^B}+\frac{\partial H}{\partial
p_B}\beta_{iA}\Phi^i_B-\frac{\partial H}{\partial
p_B}\beta_{iB}\Phi^i_A
\\
&=& \frac{\partial H}{\partial q^A}+\dot{\sigma}^B(t)
\frac{\partial \gamma_A}{\partial q^B}-\frac{\partial H}{\partial
p_B}\beta_{iB}\Phi^i_A \\
&=&-\left(\bar{\lambda}_i +\frac{\partial H}{\partial
p_B}\beta_{iB}\right) \Phi^i_A \ \hbox{(from (\ref{unotres}))}\\
\end{eqnarray*}
Therefore, we conclude that $d(H\circ \gamma)\in D^0$.

$(\Longleftarrow)$ Assume that (ii) holds, that is,
\[
\left[\frac{\partial H}{\partial q^A}+\frac{\partial H}{\partial
p_B}\frac{\partial \gamma_B}{\partial q^A}\right]\,
dq^A=\tilde{\lambda}_i {\mu}^i
\]
Now using (\ref{uno}) and since $d\gamma\in {\mathcal I} (D^0)$,
then
\[
\frac{\partial H}{\partial
q^A}+\dot{\sigma}^B(t)\left(\frac{\partial \gamma_A}{\partial
q^B}-\beta_{iA}\Phi^i_B+\beta_{iB}\Phi^i_A\right)=
\tilde{\lambda}_i \Phi^i_A
\]
Therefore
\begin{equation}\label{aqw}
\frac{d}{dt}(\gamma_A(\sigma(t))) = - \frac{\partial H}{\partial
q^A}(\gamma(\sigma(t))) +\left(
\tilde{\lambda}_i-\dot{\sigma}^B(t)\beta_{iB}(\sigma(t))\right)
{\Phi^i_A}(\sigma(t))
\end{equation}

Using that $\hbox{Im}(\gamma)\in \bar{M}$, we deduce that
$\bar{\lambda}_i=\frac{\partial H}{\partial
p_B}\beta_{iB}-\tilde{\lambda}_i$ along $\gamma$.
\end{proof}

\begin{remark}
{\rm Suppose that $\gamma=dS$ where $S$ is a function $S:
Q\longrightarrow \R$. In this case, the condition $d\gamma\in
{\mathcal I} (D^0)$ is trivially satisfied. Moreover, we note that
in previous approximations to Hamilton-Jacobi theory
\cite{Eden,Doo1,Pa,Rumyantsev,Sumbatov} the considered sections
are of the form
\begin{equation}\label{zse} \gamma(q)=(q, \frac {\partial S}{\partial
q^A}-\tilde{\lambda}_i\mu^i_A), \end{equation} and the
coefficients $\tilde{\lambda}_i$ are determined through the
nonholonomic constraint equations
\[
\mu^i_A(q)\frac{\partial H}{\partial p_A}(q, \gamma_A(q))=0\; .
\]
In general, this type of 1-forms does not satisfy the condition
that we initially impose, $d\gamma\in {\mathcal I} (D^0)$. Observe
that in the particular case of holonomic constraints both
approaches coincide.}
\end{remark}

Now, we write a coordinate expression for the Hamilton-Jacobi
equation that we have proposed. In order to do it, consider a set
of independent vector fields $\{Z_a=Z_a^A\frac{\partial}{\partial
{q}^A}\}$, $1\leq a\leq n-m$, on $Q$ such that $\mu^i(Z_a)=0$,
i.e, $D_q=\hbox{span }\{(Z_a)|_q\}$. Thus a 1-form $\gamma$ on
$Q$, solution of the nonholonomic Hamilton-Jacobi equation, must
verify the condition $d\gamma\in {\mathcal I} (D^0)$ and,
additionally,
\begin{eqnarray*}
Z_a^A(q)\left( \frac{\partial H}{\partial q^A}(q,
\gamma(q))+\frac{\partial H}{\partial p_B}(q,
\gamma(q))\frac{\partial
\gamma_B}{\partial q^A}(q)\right)&=&0, \\
\mu^i_A(q)\frac{\partial H}{\partial p_A}(q, \gamma(q))&=&0,
\end{eqnarray*}
for the condition $d(H\circ \gamma)\in D^0$ and for the condition
$\gamma(Q) \subset \bar{M}$, correspondingly.

\begin{theorem}\label{nhhj2}
Let $X$ be vector field on $Q$ such that $X(Q) \subset M$ and
$d(FL \circ X)\in {\mathcal I} (D^0)$. Then the following
conditions are equivalent:

\begin{itemize}
\item[(i)] for every curve $\sigma: \R \longrightarrow Q$ such that
\begin{equation}\label{unoo}
\dot{\sigma}(t) = T\tau_Q(X_{nh}(X(\sigma(t))))
\end{equation}
for all $t$, then $X \circ \sigma$ is an integral curve of
$X_{nh}$.

\item[(ii)] $d(E_L \circ X) \in D^0$.
\end{itemize}
\end{theorem}

\begin{definition}
A vector field $X$ satisfying the conditions of Theorem \ref{nhhj1}
will be called a solution for the Hamilton-Jacobi problem given by
$L$ and $M$.
\end{definition}

\subsection{An application to \v{C}aplygin systems}

Consider now the case of a \v{C}aplygin system (see Section
\ref{chaply}). That is, we have a fibration $\rho : Q
\longrightarrow N$, and an Ehresmann connection $\Gamma$ in
$\rho$, whose horizontal distribution imposes the constraints to a
lagrangian $L : TQ \longrightarrow \R$.

Let $L^* : TN \longrightarrow \R$ be the reduced lagrangian and
$\alpha^*$ the corresponding external force. We denote by $X_{nh}$
the nonholonomic vector field on $TQ$ and by $X^*$ the solution of
the reduced lagrangian system with external force $\alpha^*$.

\begin{theorem}
\begin{itemize}

\item[(i)] Assume that a vector field $X$ on $Q$ is a solution for the
Hamilton-Jacobi problem given by $L$ and $\Gamma$. If $X$ is
$\rho$-projectable to a vector field $Y$ on $N$ and $d(FL^*\circ
Y)=0$ then $Y$ is a solution of the Hamilton-Jacobi problem given
by $L^*$ and $\alpha^*$.

\item[(ii)] Conversely, let $Y$ be a vector field which is a
solution of the Hamilton-Jacobi problem given by $L^*$ and
$\alpha^*$. Then, if $d(FL \circ Y^{{{\mathcal H}}})\in {\mathcal
I} ({\mathcal H}^0)$, the horizontal lift $Y^{{{\mathcal H}}}$ is
a solution for the Hamilton-Jacobi problem given by $L$ and
$\rho$.
\end{itemize}

\end{theorem}

\begin{proof}

$(\Longrightarrow)$

Assume that a vector field $X$ on $Q$ is a solution for the
Hamilton-Jacobi problem given by $L$ and $\Gamma$, and that $X$ is
$\rho$-projectable onto a vector field $Y$ on $N$. We have to
prove that $Y$ is then a solution of the Hamilton problem given
$L^*$ and $\alpha^*$. Let $\mu$ a curve in $N$ such that
\begin{equation}\label{unoo2}
\dot{\mu}(t) = T\tau_N(Y^*(Y(\mu(t))))
\end{equation}
for all $t$. Take an horizontal lift $\sigma$ of $\mu$ to $Q$ with
respect to the connection $\Gamma$. A direct computation shows
that
\begin{equation}\label{unoo3}
\dot{\sigma}(t) = T\tau_Q(X_{nh}(X(\sigma(t))))
\end{equation}
since $X_{nh}$ is the horizontal lift of $Y^*$ with respect to the
prolongated connection $\bar{\Gamma}$. Therefore we have that $X
\circ \sigma$ is an integral curve of $X_{nh}$ and, consequently,
$Y \circ \mu$ is an integral curve of $Y^*$.

$(\Longleftarrow)$

Assume that $Y$ is vector field on $N$ which is a solution of the
Hamilton-Jacobi problem given by $L^*$ and $\alpha^*$. Take its
horizontal lift $X=Y^{\mathcal H}$ to $Q$. with respect to
$\Gamma$. If $\sigma$ is a curve in $Q$ satisfying
\begin{equation}\label{unoo4}
\dot{\sigma}(t) = T\tau_Q(X_{nh}(X(\sigma(t))))
\end{equation}
then the projection $\mu = \rho \circ \sigma$ satisfies
(\ref{unoo2}). So, $Y \circ \mu$ is an integral curve of $Y^*$ and,
hence $X \circ \sigma$ is an integral curve of $X_{nh}$.

\end{proof}
\begin{example} (The mobile robot with fixed orientation revisited)
{\rm The reduced lagrangian in this case is
\[
L^*(\theta,
\psi)=\frac{1}{2}J\dot{\theta}^2+\frac{mR^2+3J_{\omega}}{2}\dot{\psi}^2\;
,
\]
and $\alpha^*=0$

Therefore, \[ Y_1=\frac{\partial}{\partial \theta} \qquad
\hbox{and}\qquad Y_2=\frac{\partial}{\partial \psi}
\]
are solutions of the Hamilton-Jacobi problem given by $(L^*,
\alpha^*)$. Calculating the horizontal lifts off both vector
fields we have that:
\[
Y_1^{{{\mathcal H}}}=\frac{\partial}{\partial \theta}\qquad
\hbox{and}\qquad Y_2^{{{\mathcal H}}}=\frac{\partial}{\partial
\psi}+R\cos\theta\frac{\partial}{\partial x}+R\sin
\theta\frac{\partial}{\partial y}\; .
\]
Now
\begin{eqnarray*}
\gamma_1&=&FL\circ Y_1^{{{\mathcal H}}}=J\,d\theta\\
\gamma_2&=&FL\circ Y_2^{{{\mathcal
H}}}=3J_{\omega}\,d\psi+mR\cos\theta\, dx+mR\sin \theta\, dy
\end{eqnarray*}
and
\begin{eqnarray*}
d\gamma_1&=&0\in {\mathcal I}
({\mathcal H}^0)\\
d\gamma_2&=&-mRd\theta\wedge (\sin\theta\,dx-\cos\theta\, dy)\in
{\mathcal I} ({\mathcal H}^0)
\end{eqnarray*}
Therefore, $Y_1^{{{\mathcal H}}}$ and $Y_2^{{{\mathcal H}}}$ are
solutions of the Hamilton-Jacobi problem of the nonholonomic
problem given by $(L, {\mathcal H})$. Observe that in both cases
$d(H\circ\gamma_i)=0$, for $i=1,2$. In such a case,
\begin{eqnarray*}
t&\longmapsto& (x_0, y_0, t+\theta_0, \psi_0)\\
t&\longmapsto& (tR\cos\theta_0+x_0, tR\sin\theta_0+y_0, \theta_0,
t+\psi_0)
\end{eqnarray*} are the solutions of the nonholonomic
system $(L, {\mathcal H})$ obtained from $Y_1^{{{\mathcal H}}}$
and $Y_2^{{{\mathcal H}}}$, respectively. Both solutions are also
solutions of the lagrangian system determined by $L$ without
constraints; indeed, they are solutions of the free system
satisfying additionally the nonholonomic constraints.

But taking now the vector field
\[ Y_3=Y_1+Y_2=
\frac{\partial}{\partial \theta} + \frac{\partial}{\partial \psi}
\]
it is obviously a solution of the Hamilton-Jacobi equations for
the lagrangian $L^*$ and its horizontal lift
\[
Y_3^{{{\mathcal H}}}=\frac{\partial}{\partial
\theta}+\frac{\partial}{\partial
\psi}+R\cos\theta\frac{\partial}{\partial x}+R\sin
\theta\frac{\partial}{\partial y}\;
\]
is a solution of the Hamilton-Jacobi equations for the
nonholonomic system $(L, H)$:
\begin{eqnarray*}
\gamma_3&=&FL\circ Y_3^{{{\mathcal
H}}}=J\,d\theta+3J_{\omega}\,d\psi+mR\cos\theta\, dx+mR\sin
\theta\, dy
\end{eqnarray*}
and $ d\gamma_3\in {\mathcal I} ({\mathcal H}^0)$. In such a case,
the solution of the nonholonomic problem that we obtain is
\begin{eqnarray*}
t&\longmapsto& (R\sin (t-\theta_0)+x_0+R\sin\theta_0, -R\cos(t-\theta_0)
+y_0+R\cos\theta_0, t+\theta_0, t+\psi_0) \end{eqnarray*}
which is a solution of the nonholonomic problem but not of the free system.

}
\end{example}

\section*{Acknowledgments}
The authors would like to thank the referees for the careful
reading of the paper and the interesting remarks, which have
highly improved the content of the paper.

\end{document}